\documentclass{article}

\usepackage[english]{babel}

\usepackage[letterpaper,top=3cm,bottom=3cm,left=3cm,right=3cm,marginparwidth=1.75cm]{geometry}

\usepackage{setspace}
\usepackage{amsmath}
\usepackage{multirow}
\usepackage{amsfonts}
\usepackage{booktabs}
\usepackage{graphicx}
\usepackage{bm}
\usepackage[colorlinks=true, allcolors=blue]{hyperref}
\usepackage[round]{natbib}
\title{A note on auxiliary mixture sampling for Bayesian Poisson models}
\author{Aldo Gardini\thanks{Department of Statistical Sciences, University of Bologna, Italy}, Fedele Greco\footnotemark[1], Carlo Trivisano\footnotemark[1]}
\date{}
\begin{document}
\maketitle

\begin{abstract}
\noindent Bayesian hierarchical Poisson models are an essential tool for analyzing count data. However, designing efficient algorithms to sample from the posterior distribution of the target parameters remains a challenging task for this class of models. Auxiliary mixture sampling algorithms have been proposed to address this issue. They involve two steps of data augmentations: the first leverages the theory of Poisson processes, and the second approximates the residual distribution of the resulting model through a mixture of Gaussian distributions. In this way, an approximated Gibbs sampler is obtained. In this paper, we focus on the accuracy of the approximation step, highlighting scenarios where the mixture fails to accurately represent the true underlying distribution, leading to a lack of convergence in the algorithm. 
We outline key features to monitor, in order to assess if the approximation performs as intended. Building on this, we propose a robust version of the auxiliary mixture sampling algorithm, which can detect approximation failures and incorporate a Metropolis-Hastings step when necessary. Finally, we evaluate the proposed algorithm together with the original mixture sampling algorithms on both simulated and real datasets.
\end{abstract}

\section{Introduction}

In this paper, we consider Monte Carlo Markov Chain \citep[MCMC,][]{robert1999monte} sampling of Bayesian hierarchical models with Poisson likelihood. In these models, the expected value of the response variable, which is constituted by a vector of observed counts $\mathbf{y}=(y_1,\dots,y_i,\dots,y_n)^\top\in \mathbb{N}_0^n$,  is modelled as 
\begin{equation}
	\label{eq:lik_pois}
    y_i|\lambda_i\stackrel{ind}{\sim}\mathcal{P}(t_i\lambda_i),\quad i=1,\dots,n;
\end{equation}
where $t_i$ represents an offset and the intensity parameter $\lambda_i$ is linked to a linear predictor $\eta_i$ by the canonical link function $\lambda_i=\exp(\eta_i)$, $i=1,\ldots,n$. The primary focus of the paper is on a widespread subclass of hierarchical models known as Latent Gaussian Models (LGMs). They are characterized by a linear predictor $\boldsymbol{\eta}=(\eta_1, \ldots, \eta_i,\ldots,\eta_n)^\top$ constituted by a priori independent additive components distributed as Gaussian random variables conditionally on model hyperparameters:
\begin{equation}\label{eq:linpred}
\bm\eta=\mathbf{X}\boldsymbol{\beta}+\sum_{q=1}^Q\textbf{Z}_q\boldsymbol{\gamma}_q.
\end{equation}
The design matrix $\mathbf{X}\in\mathbb{R}^{n\times (P+1)}$ is associated to an overall intercept $\beta_0$ and $P$ fixed effects collected in the vector  $\boldsymbol{\beta}=(\beta_0,\beta_1,\ldots,\beta_P)^\top$ . Random components are expressed as the product of a random effect design matrix $\textbf{Z}_q\in\mathbb{R}^{n\times m_q}$, with $m_q\leq n$ and a random vector $\boldsymbol{\gamma}_q\in\mathbb{R}^{m_q}$, $q=1,\ldots,Q$, which in LGMs follows a Gaussian distribution. 

In a Bayesian framework, specification of the Poisson regression mixed model outlined in equations \eqref{eq:lik_pois}-\eqref{eq:linpred} can be completed by a Gaussian prior on the coefficients vector is $\boldsymbol{\beta}\sim\mathcal{N}_{P+1}(\boldsymbol{0}, \mathbf{V}_0)$, while the prior for the random effects is $\boldsymbol{\gamma}_q|\sigma_q^2\sim\mathcal{N}_{m_q}\left(\boldsymbol{0}, \sigma_q^2\mathbf{S}_q\right)$, where $\sigma_q^2$ is a scale parameter and $\mathbf{S}_q$ is a possibly structured covariance matrix. Model hierarchy is completed by prior specification for scale parameters $\sigma_q^2$, however the key points of the algorithms presented in this paper are not strictly related to this prior choice.

The problem of estimating Poisson regression models in the Bayesian paradigm has motivated several research works in the field of computational statistics, to propose efficient MCMC algorithms. Indeed, due to the lack of conditional conjugacy under any prior setting, the use of Metropolis-Hastings or Hamiltonian Monte Carlo algorithms is required, and some strategies for the Poisson model have been proposed for particular contexts (e.g. \cite{knorr2002block} focused on spatial models). 

Alternatively, a popular strand of literature focuses on the adoption of data augmentation schemes, pioneered by  \citep{albert1993bayesian} in the framework of logistic regression. In the context of Poisson models, extremely appealing proposals are those by \citet{fruhwirth2006auxiliary} and \citet{fruhwirth2009improved}. The authors proposed a two-step data augmentation approach: the first step reformulates the model in terms of inter-arrival times, leveraging the theory of Poisson processes, while the second step approximates the residuals of the augmented model using a mixture of Gaussian distributions. This approach is particularly appealing for estimating Poisson LGMs, as it facilitates the adaptation of efficient MCMC samplers designed for linear LGMs to the Poisson case. Specifically, these strategies enable the use of Gibbs samplers, as the full conditionals of the coefficients and random effects in LGMs reduce to Gaussian distributions. Consequently, the algorithms are straightforward to implement, and many R packages, such as \texttt{MCMCpack} \citep{MCMCpack} and \texttt{pogit} \citep{dvorzak2016sparse}, utilize these methods to fit Poisson models. From a different perspective, an approximation-based approach was also explored by \citet{d2023efficient}. This work exploited the convergence of the Negative Binomial distribution to the Poisson distribution to incorporate the Polya-Gamma augmentation scheme by \citet{polson2013bayesian}, using it to construct a proposal distribution within a Metropolis-Hastings algorithm.

In this paper we focus on the data augmentation schemes proposed in \cite{fruhwirth2006auxiliary} and \cite{fruhwirth2009improved}. We show that a careful check of the approximation accuracy is required, as lack of convergence can be observed in some situations where approximation's accuracy deteriorates. For this reason we propose an adjusted mixture approximation and implement an algorithm that can automatically detect cases where a Metropolis Hastings step needs to be added to the MCMC algorithm in order to guarantee convergence. Our aim is to adopt the mixture approximation proposed in \cite{fruhwirth2009improved} as a best choice, because of its computational convenience, and to resort to MH only when needed. We demonstrate the usefulness of our proposal both on simulated data and on a real case study. The proposed algorithm is implemented in the R package \texttt{SamplerPoisson} and the R code for reproducing all the computations presented in the paper is available as supplementary material. 

The remainder of the paper is structured as follows. Section \ref{sec:schemes} reviews the two auxiliary mixture sampling algorithms currently available in the literature. Section \ref{sec:pitfalls} highlights issues related to the accuracy of the mixture approximation step, while Section \ref{sec:proposal} introduces a robust version of the sampling algorithms designed to address these approximation challenges. Section \ref{sec:applications} compares the performance of the discussed algorithms on both simulated and real data. Finally, Section \ref{sec:conclusion} provides concluding remarks.

\section{Augmentation schemes}\label{sec:schemes}

Exploiting the properties of the Poisson process, \citet{fruhwirth2006auxiliary} expressed the observations in a Poisson regression model as a sequence of inter-arrival times, leading to a data augmentation scheme that is described in Section \ref{sec:2006_scheme}. This algorithm was improved by \citet{fruhwirth2009improved}, leading to an extremely appealing and computationally convenient algorithm that we summarise in Section \ref{sec:2009_scheme}.

\subsection{Auxiliary mixture sampling}\label{sec:2006_scheme}
To express the Poisson distribution in terms of inter-arrival times, \citet{fruhwirth2006auxiliary} introduce $y_i+1$ independent Exponential random variables with intensity parameter $\lambda_i$ for each sampling count $y_i$, $i=1,\dots,n$. This sequence of random variables is defined as
\begin{equation}
	\label{eq:tau_1}
	\tau_{ij}|\lambda_i=\frac{\zeta_{ij}}{\lambda_i};\ j=1,\dots,y_i+1;\  i=1,\dots,n;
\end{equation}
where $\zeta_{ij}\stackrel{ind}{\sim}\text{Exp}(1)$, $\forall (i,j)$. Taking the negative of the logarithm, the model can be expressed through the following linear relationship:
\begin{equation}
    \label{eq:reg_logtau}
y^\ast_{ij}=\mathbf{x}_i^\top\boldsymbol{\beta}+\sum_{q=1}^Q\mathbf{z}_{qi}^\top\boldsymbol{\gamma}_q+\varepsilon_{ij},\ \forall (i,j);
\end{equation}
where $ y^\ast_{ij}=-\log(\tau_{ij})$ is the auxiliary variable determining the linear model and the error term $\varepsilon_{ij}=-\log\left(\zeta_{ij}\right)$ follows a Negative Log-Gamma (NLG) distribution. Such distribution plays a fundamental role in both augmentation schemes: the density function of a random variable $U\sim\text{NLG}(\psi,1)$ is
\begin{equation}
	\label{eq:nlg_dens}
	f_U(u)=\frac{\exp\left\{-u\psi-e^{-u}\right\}}{\Gamma(\psi)}.
\end{equation}
Under this augmentation scheme, the $n^\ast = n + \sum_{i=1}^ny_i$ auxiliary variables follow a NLG distribution with $\psi=1$: 
$$\varepsilon_{ij}\stackrel{ind}{\sim}\text{NLG}(1,1),\qquad\forall (i,j).$$ 

\cite{fruhwirth2006auxiliary} proposed to introduce a second step of data augmentation by approximating the density function \eqref{eq:nlg_dens} with $\psi=1$ through a mixture of $K$ Gaussian distributions
\begin{equation}
    \label{eq:mixt-app}
    g_{\varepsilon}(z)=\sum_{k=1}^K w_k\phi\left(z;m_k,s^2_k\right).  
\end{equation}
In this expression, $\phi(z;m_k,s_k^2)$ indicates the density function of the $k$-th Gaussian component, evaluated in $z$ with mean $m_k$, variance $s_k^2$, and having weight $w_k$. Under this approximation for the distribution of the error term, a Gaussian linear model emerges conditionally on the labels $r_{ij}\in\{1,\dots,K\}$ which identify the mixture component:
$$
y^\ast_{ij}|r_{ij}=\mathbf{x}_i^\top\boldsymbol{\beta}+\sum_{q=1}^Q\mathbf{z}_{qi}^\top\boldsymbol{\gamma}_q+m_{r_{ij}}+\varepsilon_{ij},\qquad \varepsilon_{ij}|r_{ij}\sim \mathcal{N}(0, s^2_{r_{ij}}).
$$
To express the model in matrix form, let $\mathbf{r}\in\{1,\dots,K\}^{n^\ast}$ denote the $n^\ast$-dimensional vector of mixture labels, and define the vectors containing the means and variances of the mixture components conditionally on the specific labels $\mathbf{m}(\mathbf{r})=(m_{r_{11}},\dots,m_{r_{n,y_n+1}})^\top$ and $\mathbf{s}^2(\mathbf{r})=(s^2_{r_{11}},\dots,s^2_{r_{n,y_n+1}})^\top$. Storing the auxiliary variables into $\mathbf{y}^\ast$ and the correspondent design matrices in $\mathbf{X}^\ast\in\mathbb{R}^{n^\ast\times P+1}$ and $\mathbf{Z}_q^\ast\in\mathbb{R}^{n^\ast\times M_q},\ \forall q$, the linear model conditional on the latent labels indicating the mixture components can be expressed in matrix form as:
\[
\mathbf{y}^\ast|\mathbf{r} = \mathbf{X}^\ast\boldsymbol{\beta}+\sum_{q=1}^Q\mathbf{Z}_{q}^*\boldsymbol{\gamma}_q+\mathbf{m}(\mathbf{r})+\boldsymbol{\varepsilon},\qquad {\boldsymbol{\varepsilon}}|\mathbf{r}\sim\mathcal{N}_{n^\ast}\left(\boldsymbol{0},\mathbf{D}(\mathbf{r})\right),
\]
where $\mathbf{D}(\mathbf{r})=\text{diag}\left(\mathbf{s}^2(\mathbf{r})\right)$. 

The MCMC algorithm, denoted as the AMS (Auxiliary Mixture Sampling) algorithm in what follows, is implemented by iterating through the following steps.
\begin{itemize}
    \item \textit{Step 1.} For each $i$, $y_i+1$ auxiliary variables $\tau_{ij}$, $j=1,\ldots,y_i+1$ are generated.
    \begin{itemize}
    	\item For $j=1,\ldots y_i$, auxiliary variables are constituted by a sorted sample of size $y_i$ drawn from a Uniform distribution on the interval $[0,1]$ since the arrival times of a Poisson process conditioned on having observed a given number of jumps are distributed as the order statistics of a $\mathcal{U}(0,1)$ distribution.
    	\item The $(y_i+1)-$th auxiliary variable is drawn exploiting that $\tau_{i(y_i+1)}=1-\sum_{j=1}^{y_i}\tau_{ij}+\zeta_{i}/\lambda_i$, where $\zeta_{i}\sim\text{Exp}(1)$. In this way, a realization of the $n^*$ auxiliary variables $\mathbf{y}^{*}$ is drawn.
    \end{itemize}
    \item \textit{Step 2. } The mixture component indicators $r_{ij}$ are sampled with probabilities proportional to $\mathbb{P}[r_{ij}=k|y_{ij}^{*},\boldsymbol{\beta},\boldsymbol{\gamma}_1,\ldots,\boldsymbol{\gamma}_Q]$, obtaining $\mathbf{r}$.
    \item \textit{Step 3.} Draw $\boldsymbol\beta$ from the approximated full conditional:
\begin{equation}\label{eq:FC_beta}
	\boldsymbol\beta|\mathbf{y}^{\ast},\mathbf{r},\boldsymbol{\gamma}_1,\ldots,\boldsymbol{\gamma}_Q\sim N\left(\boldsymbol{\mu}_\beta\left(\mathbf{r}\right), \boldsymbol{\Sigma}_\beta\left(\mathbf{r}\right)\right),
\end{equation}
where 
$$
\boldsymbol{\Sigma}_\beta\left(\mathbf{r}\right)=\left(\mathbf{X}^{\ast^\top}\mathbf{D}\left(\mathbf{r}\right)^{-1}\mathbf{X}^\ast+\mathbf{V}_0^{-1}\right)^{-1},
$$ and $$\boldsymbol{\mu}_\beta\left(\mathbf{r}\right)=\boldsymbol{\Sigma}_\beta\mathbf{X}^{\ast^\top}\mathbf{D}\left(\mathbf{r}\right)^{-1}\left(\mathbf{y}^{\ast}-\sum_{q=1}^Q\mathbf{Z}_{q}^*\boldsymbol{\gamma}_q-\mathbf{m}\left(\mathbf{r}\right)\right).$$
\item \textit{Step 4.} Draw $\boldsymbol{\gamma}_q,\ \forall q,$ from the approximated full conditional distributions: 

\begin{equation}\label{eq:FC_gamma}
	\boldsymbol\gamma_q|\mathbf{y}^{\ast},\mathbf{r},\boldsymbol{\beta}, \sigma^{2}_{\gamma_q}\sim N\left(\boldsymbol{\mu}_{\gamma_q}\left(\mathbf{r}\right), \boldsymbol{\Sigma}_{\gamma_q}\left(\mathbf{r}\right)\right),
\end{equation}
where 
$$
\boldsymbol{\Sigma}_{\gamma_q}\left(\mathbf{r}\right)=\left(\mathbf{Z}_q^{\ast^\top}\mathbf{D}\left(\mathbf{r}\right)^{-1}\mathbf{Z}_q^\ast+\sigma^{-2}_{\gamma_q}\mathbf{K}_{q}\right)^{-1},
$$ and $$\boldsymbol{\mu}_{\gamma_q}\left(\mathbf{r}\right)=\boldsymbol{\Sigma}_{\gamma_q}\mathbf{Z}_q^{\ast^\top}\mathbf{D}\left(\mathbf{r}\right)^{-1}\left(\mathbf{y}^{\ast}-\mathbf{X}^{\ast}\boldsymbol\beta-\sum_{j\neq q}\mathbf{Z}_{j}^*\boldsymbol{\gamma}_j-\mathbf{m}\left(\mathbf{r}\right)\right).$$
\item \textit{Step 5.} Draw $\sigma_q^{2},\ \forall q,$ from the full conditional or performing an MH step.
\end{itemize}
Note that Steps 3-5 are standard steps for sampling a LGM with Gaussian likelihood, thanks to the auxiliary variables introduced in Steps 1-2. 

\subsection{Improved auxiliary mixture sampling}\label{sec:2009_scheme}
The main drawback of the AMS algorithm is the large number of latent variables that are introduced, particularly when the overall number of observed counts $\sum_{i=1}^ny_i$ is huge. To address this issue, \cite{fruhwirth2009improved} proposed a more parsimonious scheme, requiring for each observation two auxiliary variables if $y_i > 0$, and one auxiliary variable if $y_i = 0$. Denoting the number of observations equal to zero as $n_0$, the total number of latent variables is $\widetilde n^* = 2n - n_0$, which can be much smaller than $n^\ast$.

In this improved scheme, a Poisson process is defined over the time interval $t\in[0, 1]$. For each observation $i$, two latent variables are defined conditionally on $y_i > 0$ jumps occurring before $t = 1$. The variable $\widetilde\tau_{i2}$ represents the time of arrival after $y_i$ jumps before $t = 1$, while $\widetilde\tau_{i1}$ represents the inter-arrival time of the first jump after $t = 1$. Consequently, the sum $\widetilde\tau_{i2} + \widetilde\tau_{i1}$ gives the arrival time of the first jump after $t = 1$. For observations where $y_i = 0$, $\widetilde\tau_{i2}$ is known to be $0$, so only the latent variable $\widetilde\tau_{i1}$ is required.

The set of latent variables $\widetilde\tau_{i1}$ is defined as in Equation \eqref{eq:tau_1}, while the distribution of $\widetilde\tau_{i2}$ follows from the fact that the arrival time is the sum of $y_i$ independent Exponential distributions with rate $\lambda_i$, hence:
$$
\widetilde\tau_{i2}=\frac{\widetilde\zeta_{i2}}{\lambda_i}, \quad \widetilde\zeta_{i2}\sim\text{Gamma}(y_i,1),\ \forall i.
$$

Likewise the AMS data augmentation scheme, it is possible to express the model through a linear relationship as Equation \eqref{eq:reg_logtau}:
\begin{equation}
    \label{eq:reg_logtauprime}
\widetilde y^\ast_{ij}=\mathbf{x}_i^\top\boldsymbol{\beta}+\sum_{q=1}^Q\mathbf{z}_{qi}^\top\boldsymbol{\gamma}_q+\widetilde\varepsilon_{ij},
\end{equation}
where $\widetilde y^\ast_{ij}=-\log\left(\widetilde\tau_{ij}\right)$ and $\widetilde\varepsilon_{ij}=-\log\left(\widetilde\zeta_{ij}\right)$, with $j=1$ if $y_i=0$ and $j=\{1,2\}$ if $y_i>0$. Concerning the error term, $\widetilde\varepsilon_{ij}\sim\text{NLG}(1,1)$ for $j=1$ while, if $y_i>0$, $\widetilde\varepsilon_{ij}\sim\text{NLG}(y_i,1)$ when $j=2$. The distribution of $\widetilde\varepsilon_{i1}$ is approximated by the same mixture of Equation \eqref{eq:mixt-app}. Conversely, the approximation of the distribution of $\widetilde\varepsilon_{i2}$ has to depends on $y_i$. \cite{fruhwirth2009improved} discussed and tabulated the different vectors of weights, means and variances that characterize the mixture approximation in Equation \eqref{eq:mixt-app} for $y\in\mathbb{N}_0$.

The sampling scheme derived from this data augmentation strategy will be referred to as the IAMS (Improved Auxiliary Mixture Sampling) algorithm. It essentially follows the same steps as the AMS algorithm, with the only difference occurring in Step 1:
\begin{itemize}
    \item \textit{Step 1.} Sample $\widetilde\zeta_{i}$ from an $\text{Exp}(1)$ random variable.
    \begin{itemize}
    	\item If $y_i=0$, set $\widetilde\tau_{i1}=1+\widetilde\zeta_i/\lambda_i$
    	\item If $y_i>0$, draw $\widetilde\tau_{i2}$ from a $\text{Beta}(y_i,1)$ and set $\widetilde\tau_{i1}=1-\widetilde\tau_{i2}+\widetilde\zeta_i/\lambda_i$.
    \end{itemize}
    \item \textit{Steps 2-5.} As AMS algorithm. 
\end{itemize}

\section{Investigating the accuracy of the mixture approximation}
\label{sec:pitfalls}
In this section, our goal is to point out some criticisms related to the second data augmentation step of AMS and IAMS algorithms, namely the approximation of the residuals distributions through a mixture of Gaussian random variables. As highlighted by \citet{fruhwirth2006auxiliary} and \citet{fruhwirth2009improved}, the accuracy of the approximation is generally excellent. However, some issues may arise in the tails, where the decay of the NLG distribution is significantly faster in the left tail and slower in the right tail compared to the Gaussian distributions used in the approximation. It is important to note that in the vast majority of applications where we tested the AMS and IAMS algorithms, the MCMC chains successfully converged. Nonetheless, there may be instances where the accuracy of the mixture approximation is insufficient to ensure convergence of the algorithms to the posterior distribution. Such cases can occur when extreme residuals emerge during the first data augmentation process, often due to factors like model misspecification or the presence of outlying observations. In any case, an MCMC algorithm should reliably converge to the target posterior regardless of model misspecifications or the presence of outliers. Moreover, it is crucial to emphasize that a simple inspection of the chains generated by the algorithm does not reveal whether the posterior samples are drawn from a stationary  distribution which does not coincide with the posterior distribution.

 \begin{figure}
     \centering
     \includegraphics[width=1\linewidth]{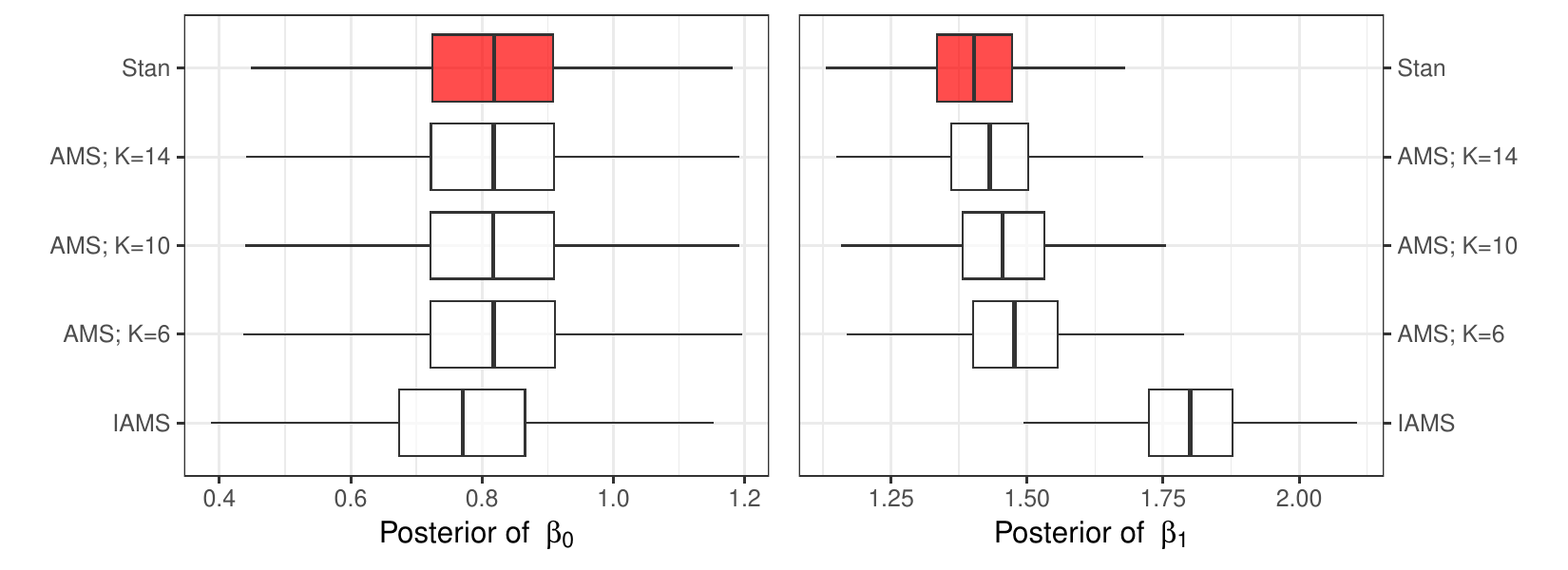}
     \caption{Boxplot of the posterior distributions for the regression coefficients $\beta_0$ and $\beta_1$.}
     \label{fig:post_bp}
 \end{figure}
 
 To better understand the issues affecting the AMS and IAMS algorithms, we consider a synthetic toy dataset with the following characteristics. A sample of size $n = 30$ is generated, with two covariates, $x_1$ and $x_2$, drawn from a standard Gaussian distribution. These covariates are then used to generate the response, which follows a Poisson distribution with log-intensity given by $\log(\lambda_i) = 0.1 + x_{1i} + 1.2x_{2i}$, $i=1,\ldots,30$. Results presented in what follows come from the estimation of the Poisson regression model with linear predictor
$$\log(\lambda_i)=\beta_0+\beta_1x_{1i},\quad i=1,\ldots,30,$$
i.e. we mimic a model misspecification setting by omitting the covariate $x_2$.

The model is firstly fitted in $\texttt{Stan}$ through its \texttt{R} interface \texttt{rstan} \citep{stan}, to get a benchmark. Then, the same model is also fitted by exploiting the AMS and IAMS algorithms. Concerning AMS, in addition to the suggested mixture approximation with $K=10$ components, we also include the mixtures with $K=6$ and $K=14$ components, to show the effect of different approximations on algorithm convergence. The mixture parameters are retrieved following the procedure outlined in Section 2.3 of \citet{fruehwirth2007auxiliary}. 

The posterior distributions of the regression coefficients, based on $B=100,000$ iterations for each algorithm, are compared in Figure \ref{fig:post_bp}, where the results obtained using \texttt{Stan} are highlighted in red to underscore that they represent the benchmark. The most noticeable discrepancies are registered using the IAMS algorithm, which leads to miscalibrated posterior distributions for both parameters. On the other hand, under the AMS algorithm the posteriors of $\beta_0$ appears to be well calibrated. Conversely, lack of convergence is observed for the posteriors of $\beta_1$. In this case, the discrepancies between the obtained posterior and the results by \texttt{Stan} gradually decrease when passing from $K=6$ to $K=14$, through the advised $K=10$ setting, highlighting the role of the approximation in these results. 

 \begin{figure}
     \centering
     \includegraphics[width=1\linewidth]{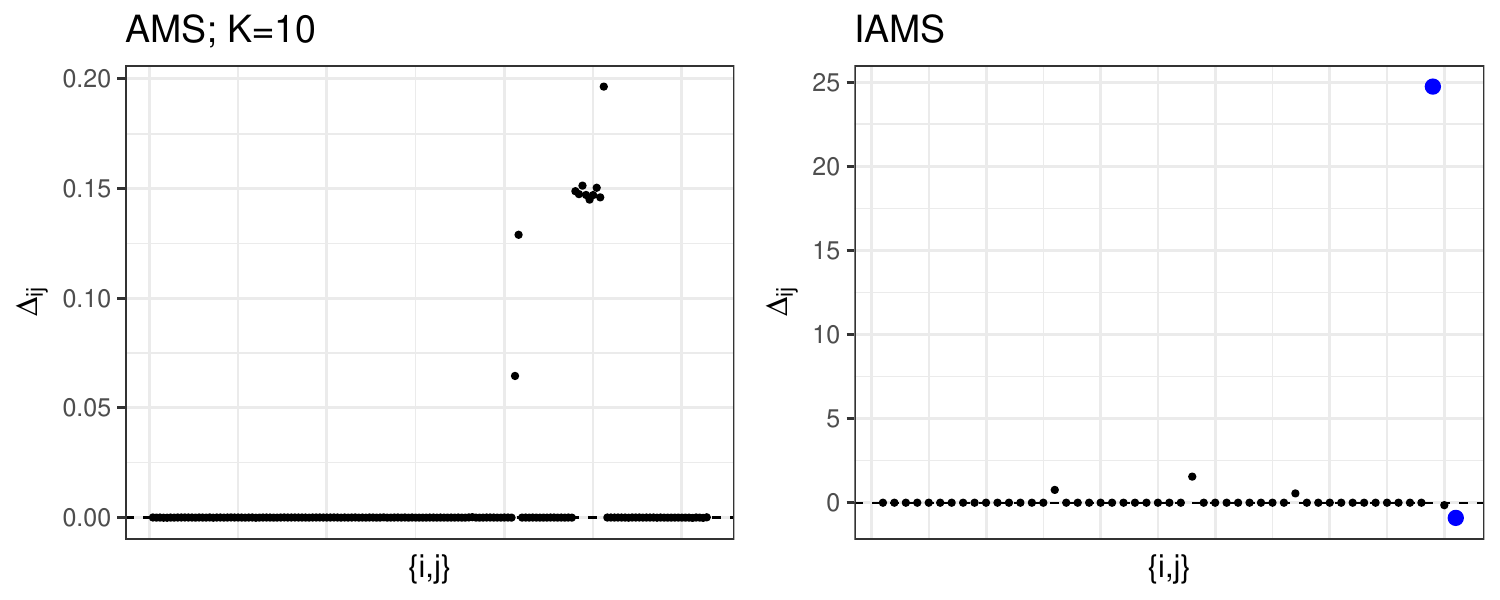}
     \caption{Average differences between the true log-likelihood and its mixture approximation. The blue dots highlight the larger values.}
     \label{fig:llik}
 \end{figure}

An insight into the reasons leading to the lack of convergence of the AMS and IAMS algorithms can be obtained by monitoring the differences between the logarithms of the true density of the residuals $\varepsilon_{ij}$ (indicated with $f_\varepsilon$) and their mixture approximation ($g_\varepsilon$). In particular, Figure \ref{fig:llik} shows the average of this discrepancy over the $B$ MCMC iterations, defined as:
\begin{equation}
    \Delta_{ij}=\frac{1}{B}\sum_{b=1}^B \left(\log g_\varepsilon \left(y_{ij}^{*(b)}-\mathbf{x}^\top_i\boldsymbol{\beta}^{(b)}\right)-\log f_\varepsilon \left(y_{ij}^{*(b)}-\mathbf{x}^\top_i\boldsymbol{\beta}^{(b)}\right)\right),
\end{equation}
where $y^{*(b)}_{ij}$ should be replaced by $\widetilde y^{*(b)}_{ij}$ when dealing with the IAMS algorithm. By comparing the results under the AMS and IAMS algorithms, an explanation for the greater robustness of AMS, as seen in the posterior distributions in Figure \ref{fig:post_bp}, becomes apparent: the values of $\Delta_{ij}$ that deviate from zero are much smaller compared to those from the IAMS algorithm.

\begin{figure}
     \centering
     \includegraphics[width=1\linewidth]{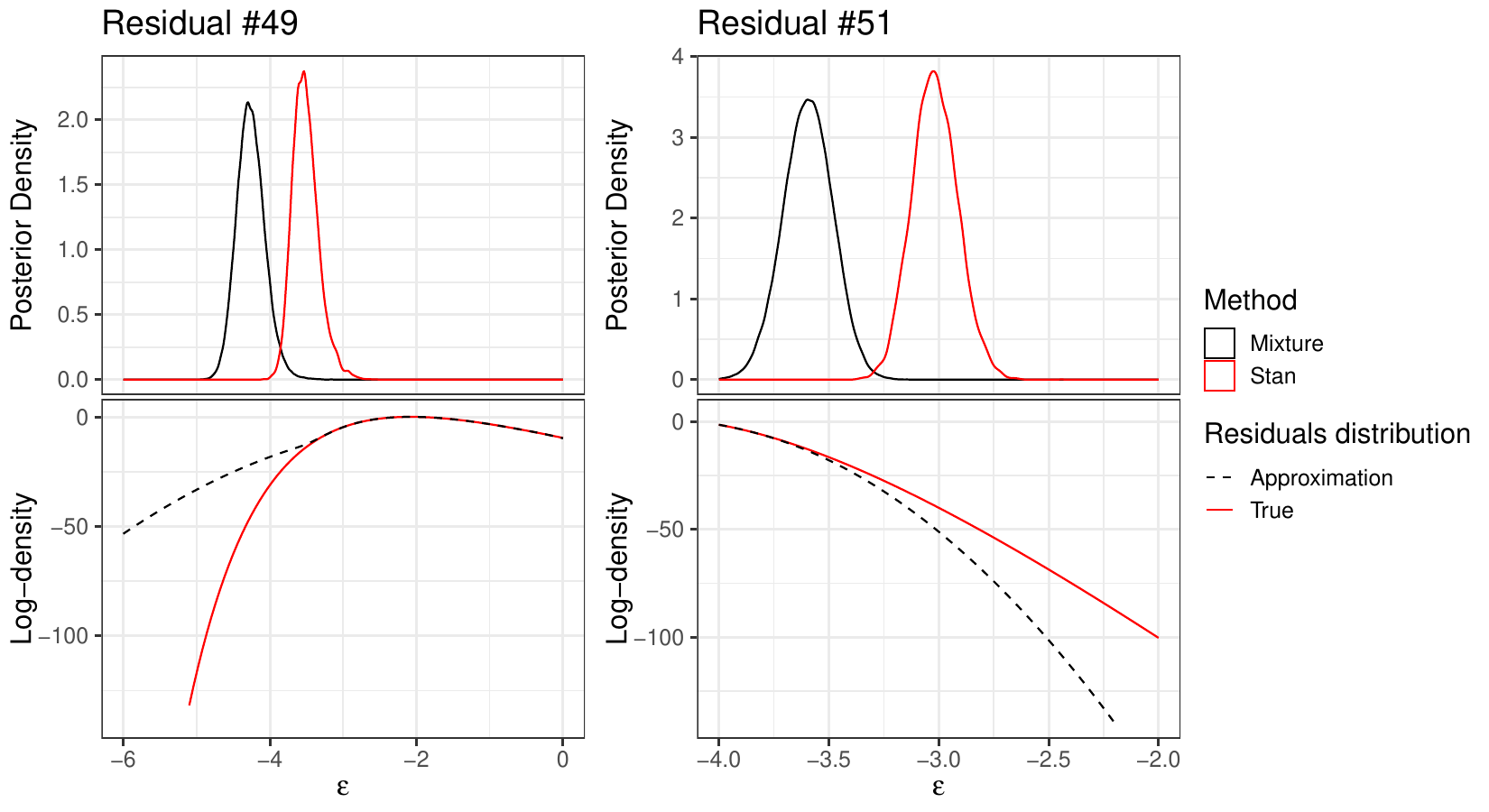}
     \caption{Upper panel plots: posterior density of residuals related to two auxiliary variables obtained using \texttt{Stan} (red) and the IAMS (black) sampler. Bottom panel plots: log density of the true auxiliary variable distribution (red) and its mixture approximation (black).}
     \label{fig:red}
 \end{figure}

The blue dots in the right panel of Figure \ref{fig:llik} refer to the latent variable residuals for which the highest positive and negative differences are observed for IAMS. Figure \ref{fig:red} focuses on these particular latent variable residuals: in the bottom line plots, the log-density of true distribution (red solid line) is compared to its mixture approximation (black dashed line), and it is possible to notice the substantial differences in the tail decay. The top-line plots report the posterior distributions of the residuals related to the selected auxiliary variables. The posteriors obtained using \texttt{Stan} (red lines) are compared to those obtained with IAMS (black lines). Negative values of $\Delta_{ij}$ are observed when the residuals are located in the right tail of the distribution (as residual \#51), and, in this situation, the approximation leads to a left shift of the posterior of the residual. For positive values of $\Delta_{ij}$, as in the case of residual \#49, a left shift is observed because the left tail of the mixture approximation is heavier than the NLG distribution: in fact, the Gaussian density cannot mimic the double exponential decay of the NLG density. 

\section{Improving convergence of the IAMS algorithm}\label{sec:proposal}
A possible way to design an MCMC algorithm that is robust with respect to potential issues caused by approximating a distribution is the inclusion of a rejection step. Indeed, as discussed in \citet{fruhwirth2006auxiliary}, the full conditional of the coefficients in Equations \eqref{eq:FC_beta} and \eqref{eq:FC_gamma} can also be exploited as proposal distributions within a Metropolis-Hastings (MH) algorithm. In this section, we first discuss some aspects related to setting an MH algorithm within the IAMS, then we propose a robust version of the IAMS algorithm.

%\subsection{Adding an MH step to the IAMS algorithm}\label{sec:MH}

To shed light on the functioning of adding an MH step to the IAMS algorithm we consider the case of sampling from the posterior of $\boldsymbol{\beta}$, but the considerations extend easily to random coefficients $\boldsymbol{\gamma}$. 
%The first aspect to notice is that the proposal distribution does not depend on past realizations and an independent Metropolis algorithm arises, conditionally on the generic vector of auxiliary variables $\mathbf{y}^*$. 
In more detail, at iteration $b$, the acceptance probability for a proposed vector of coefficients $\boldsymbol{\beta}^{\text{prop}}$ depends on the previous value $\boldsymbol{\beta}^{(b-1)}$ and the auxiliary variables $\mathbf{y}^{*(b)}$, regardless of the specific scheme under consideration. It is defined as $\alpha = \min\left(1, \pi\left(\boldsymbol{\beta}^{\text{prop}},\boldsymbol{\beta}^{(b-1)},\mathbf{y}^{*(b)}, \boldsymbol{\gamma}^{(b-1)}\right)\right)$, where 
\begin{align}
\label{eq:acc_prop_nonsimp}
    \pi\left(\boldsymbol{\beta}^{\text{prop}},\boldsymbol{\beta}^{(b-1)},\mathbf{y}^{*(b)}, \boldsymbol{\gamma}^{(b-1)}\right) &=\frac{p\left(\boldsymbol{\beta}^{\text{prop}}|\mathbf{y}^{*(b)}, \boldsymbol{\gamma}^{(b-1)}\right)}{p\left(\boldsymbol{\beta}^{(b-1)}|\mathbf{y}^{*(b)}, \boldsymbol{\gamma}^{(b-1)}\right)}\cdot \frac{\mathcal{K}\left(\boldsymbol{\beta}^{(b-1)}| \boldsymbol{\beta}^{\text{prop}}\right)}{\mathcal{K}\left(\boldsymbol{\beta}^{\text{prop}}| \boldsymbol{\beta}^{(b-1)}\right)},
\end{align}
This ratio is a function of the target posterior 
\begin{equation}
\label{eq:target}
 p\left(\boldsymbol{\beta}|\mathbf{y}^*, \boldsymbol{\gamma}\right)\propto \mathcal{L}\left(\boldsymbol{\beta}, \boldsymbol{\gamma}; \mathbf{y}^{*}\right)p(\boldsymbol{\beta})
\end{equation}
 and the transition kernel 
\begin{equation}
\label{eq:trans}
 \mathcal{K}\left(\boldsymbol{\beta}^{(b-1)}| \boldsymbol{\beta}^{\text{prop}}\right)=\mathcal{L}_a\left(\boldsymbol{\beta}^{(b-1)}, \boldsymbol{\gamma}^{(b-1)}; \mathbf{y}^{*(b)}\right)p\left(\boldsymbol{\beta}^{(b-1)}\right);
\end{equation}
which, in turn, depend on the augmented likelihood $\mathcal{L}\left(\boldsymbol{\beta}, \boldsymbol{\gamma}; \mathbf{y}^{*}\right)=\prod_{i=1}^{n}\prod_jf_\varepsilon\left(y_{ij}^*-\mathbf{x}_i^\top\boldsymbol{\beta}-\mathbf{z}_i^\top\boldsymbol{\gamma}\right)$ and its approximation  $\mathcal{L}_a\left(\boldsymbol{\beta}, \boldsymbol{\gamma}; \mathbf{y}^{*}\right)=\prod_{i=1}^{n}\prod_jg_\varepsilon\left(y_{ij}^*-\mathbf{x}_i^\top\boldsymbol{\beta}-\mathbf{z}_i^\top\boldsymbol{\gamma}\right)$. Note that the range of index $j$ is left unspecified as it depends on the data augmentation scheme.
Finally, plugging equations \eqref{eq:target} and \eqref{eq:trans} into \eqref{eq:acc_prop_nonsimp} one obtains the acceptance probability as

\begin{equation}\label{eq:accept}
    \pi\left(\boldsymbol{\beta}^{\text{prop}},\boldsymbol{\beta}^{(b-1)},\mathbf{y}^{*(b)}, \boldsymbol{\gamma}^{(b-1)}\right)=\frac{\mathcal{L}\left(\boldsymbol{\beta}^{\text{prop}}, \boldsymbol{\gamma}^{(b-1)}; \mathbf{y}^{*(b)}\right)}{\mathcal{L}\left(\boldsymbol{\beta}^{(b-1)}, \boldsymbol{\gamma}^{(b-1)}; \mathbf{y}^{*(b)}\right)}\cdot \frac{{\mathcal{L}_a}\left(\boldsymbol{\beta}^{(b-1)}, \boldsymbol{\gamma}^{(b-1)}; \mathbf{y}^{*(b)}\right)}{{\mathcal{L}_a}\left(\boldsymbol{\beta}^{\text{prop}}, \boldsymbol{\gamma}^{(b-1)}; \mathbf{y}^{*(b)}\right)}.
\end{equation}

From the expression of the transition kernel \eqref{eq:trans}, we can point out that a MH algorithm with independent proposal distribution is being considered, as $\mathcal{K}\left(\boldsymbol{\beta}^{(b-1)}| \boldsymbol{\beta}^{\text{prop}}\right)=\mathcal{K}\left(\boldsymbol{\beta}^{(b-1)}\right)$ due to the independence on $\boldsymbol{\beta}^{\text{prop}}$. In addition, the full conditional distributions of the coefficients involved in the AMS and IAMS algorithms play the role of proposal distribution in the MH sampler. 

When dealing with a MH algorithm with independent proposal, it is important to check conditions for convergence, as the ratio between the proposal and the target strongly affects the acceptance rate. According to Theorem 2.1 in \citet{mengersen1996rates}, the convergence of an independent MH algorithm requires that it must exist a number $t>0$ such that
\begin{equation}\label{eq:convergence}
    \frac{\mathcal{K}\left(\boldsymbol{\beta}\right)}{p\left(\boldsymbol{\beta}|\mathbf{y}^{*}\right)}\geq t,\ \boldsymbol{\beta}\in\mathbb{R}^p.
\end{equation}
In other words, it is required that the tails of the proposal distribution dominate the tails of the target posterior. 

When the  Gaussian mixture is used to approximate the residuals distribution, some issues could arise, from this point of view. Indeed, as noted in the bottom line plots of Figure \ref{fig:red}, the left tail of the approximation dominates the left tail of the NLG distribution, whereas its right tail is dominated by the right tail of the NLG distribution. This feature can prevent convergence of the MH algorithm. To check if the condition \eqref{eq:convergence} is fulfilled, we consider the simple model:
\begin{equation*}
    y_i|\mu\stackrel{ind}{\sim}\mathcal{P}\left(t_i\exp\{\mu\}\right),\ i=1,\dots,n,
\end{equation*}
which leads to the following model in terms of auxiliary variables, considering any augmentation scheme: 
\begin{equation*}
    y^*_{ij} = \mu+\varepsilon_{ij},\ \forall (i,j).
\end{equation*}
Under this model, the ratio in Equation \eqref{eq:convergence} can be expressed as
\begin{equation*}
    \begin{aligned}
        \frac{\mathcal{K}\left(\boldsymbol{\beta}\right)}{p\left(\boldsymbol{\beta}|\mathbf{y}^{*}\right)}=\frac{\prod_{i=1}^{n}\prod_jg_\varepsilon\left(y_{ij}^*-\mu\right)}{\prod_{i=1}^{n}\prod_jf_\varepsilon\left(y_{ij}^*-\mu\right)}.
    \end{aligned}
\end{equation*}
Plugging into the formula the expressions of the densities \eqref{eq:nlg_dens} and \eqref{eq:mixt-app}, we obtain:
\begin{equation}
    \begin{aligned}
        \frac{\mathcal{K}\left(\boldsymbol{\beta}\right)}{p\left(\boldsymbol{\beta}|\mathbf{y}^{*}\right)}&=\frac{\prod_{i=1}^{n}\prod_j\left(\sum_{k=1}^K w_k(2\pi\sigma_k)^{-1/2}\exp\left\{-\frac{1}{2\sigma^2_k}\left(y^*_{ij}-\mu\right)^2\right\}\right)}{\prod_{i=1}^{n}\prod_j\Gamma\left(y_i\right)^{-1}\exp\left\{-(y^*_{ij}-\mu)y_i-\exp\{-y^*_{ij}+\mu\}\right\}}\\
        &=\frac{\sum_{k=1}^K\left(\frac{w_k}{\sqrt{2\pi}\sigma_k}\right)^{n^*}\exp\left\{-\frac{1}{2\sigma_k^2}\left(\sum_{i,j}y^{*2}_{ij}-2\mu\sum_{i,j}y^{*}_{ij}+n^*\mu^2\right)\right\}}{\left[\prod_i\Gamma\left(y_i\right)^{-n_i^*}\right]\exp\left\{-\sum_{i,j}y^{*}_{ij}y_{i}+\mu\sum_in_i^*y_i-\sum_{i,j}\frac{e^\mu}{e^{y^{*}_{ij}}}\right\}},
    \end{aligned}
\end{equation}
where $n_i^*$ is the number of auxiliary variables associated with the $i$-th observation. From this result, we can notice that 
\begin{equation}\label{eq:limit}
    \lim_{\mu\to+\infty} \frac{\mathcal{K}\left(\boldsymbol{\beta}\right)}{p\left(\boldsymbol{\beta}|\mathbf{y}^{*}\right)} = 0,
\end{equation}
i.e. the right tail of the proposal for $\mu$ is lighter than the right tail of the target distribution, leading to the violation of the condition \eqref{eq:convergence}. This means that an MH sampler adopting the mixture approximation used in the AMS and IAMS algorithms is not guaranteed to converge because of the fast decay of the right tail. 

The above considerations lead us to design an algorithm that incorporates both a rejection step and an adjusted mixture approximation, as detailed in the next subsection. We specifically focus on the IAMS algorithm because of its computational efficiency and suitability for managing large observed counts.

\subsection{The Robust IAMS (RIAMS) algorithm}
\label{sec:mh-iams-auto}

The first step in developing a robust version of the IAMS algorithm is to find an adjusted mixture approximation of the NLG distribution that preserves the decay of the NLG density in the right tail. While, in principle, using a mixture of Gaussian distributions cannot alter the limiting behavior described by Equation \eqref{eq:limit}, it is possible to add mixture components to significantly mitigate the issue.
Since obtaining a more accurate approximation by using the algorithm proposed in in Section 2.3 of \citet{fruehwirth2007auxiliary} is infeasible for numeric problems, we implement a rough approximation that serves only as a proposal distribution in a MH algorithm.

\begin{figure}
	\centering
	\includegraphics[width=1\linewidth]{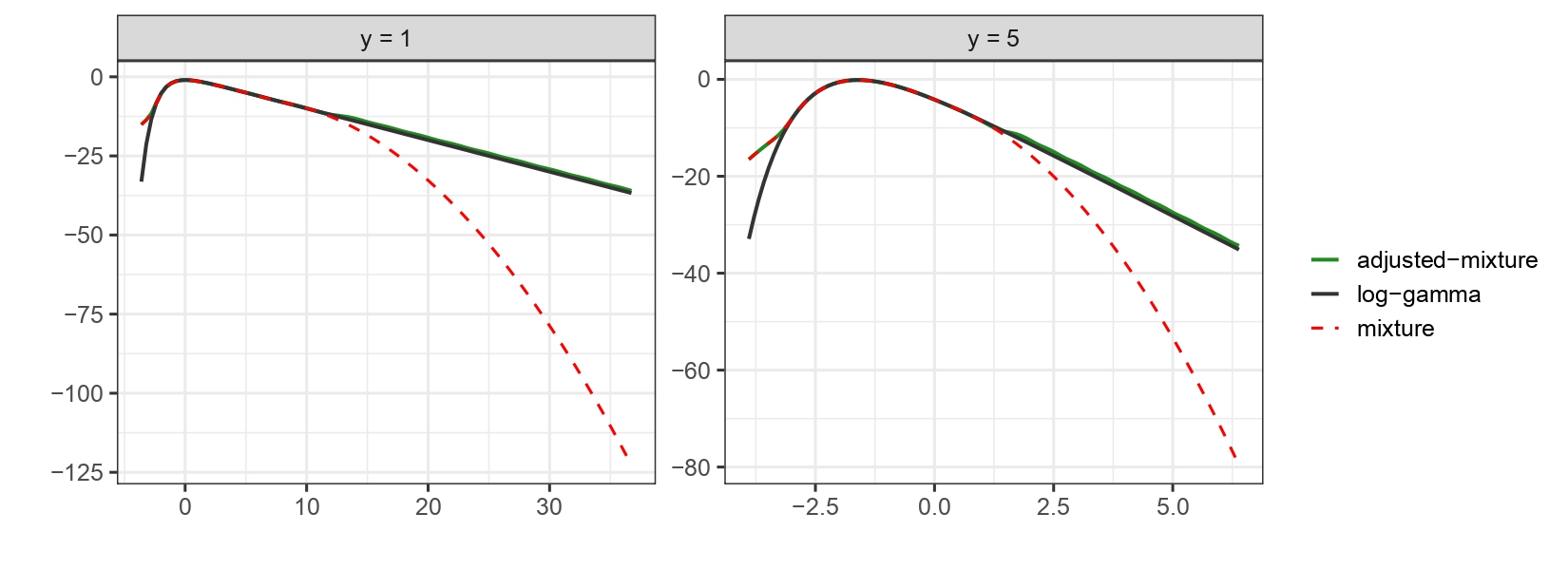}
	\caption{Mixture approximation $g_\varepsilon$ (red dashed line), adjusted mixture approximation $\widetilde g_\varepsilon$ (blue dashed line) and exact NLG log-density (black line) for $y=1$ (left panel) and $y=5$ (right panel).}
	\label{fig:ldens_adj}
\end{figure}

The adjusted approximation, denoted as ${g}_\varepsilon^\star$ in what follows, starts by identifying the point where the difference between the NLG density and the original mixture density \eqref{eq:mixt-app} become, as a rule of thumb, greater than 1 in the log scale: this point is denoted as $\xi_U(y)$ in what follows, since it depends on $y$, $y\in\mathbb{N}$. Then, we add a new mixture component centered at $m_{K+1}(y)=\xi_U(y)$, followed by 29 equally spaced knots between $m_{K+1}(y)$ and $2.5\times q_{1-10^{-16}}(y)+1.5\times \log(y)$, where $q_{1-10^{-16}}(y)$ is the quantile of order $1-10^{-16}$ of the $\text{NLG}(y,1)$  distribution. The dependence on $y$ of the right-most point where the density is approximated allows to take account of the different decay rate of the right tail of the approximated density as a function of $y$. The variance of each mixture component is computed by minimizing the difference between the true density and the mixture density at the successive knot. At each iteration, the sum of weights is normalized. Note that, as weights of the new component are very small, the accuracy of the approximation for lower quantiles is preserved, as shown in Figure \ref{fig:ldens_adj}. At the same time, the adjusted mixture more effectively captures the prolonged decay of the right tail of the NLG distribution.

Since the tail approximation is coarse, the proposed RIAMS algorithm introduces an MH step within the IAMS algorithm, leveraging the mixture approximation $\widetilde{g}_\varepsilon$ defined in Equation \eqref{eq:mixt_choice} as proposal distribution. In addition, a preliminary step is included whose aim is twofold: running some warm-up iterations with the IAMS algorithm to avoid possible issues related to initial values being far from the posterior distribution and individuating the residuals falling in the right tail, for which the adjusted mixture $g^\star_\varepsilon$ needs to be adopted. The algorithm consists of the following steps:
\begin{itemize}
	\item \textit{Step 0a.} Perform $T_1$ iterations of the IAMS algorithm. 
	\item \textit{Step 0b.}	Perform $T_2$ iterations of the IAMS algorithm and compute, for each residual, the proportions of iterations that exceed $\xi_U(y_i)$:
	\begin{equation}\label{eq:upper}
			\kappa_{ij}^U=\frac{1}{T_2}\sum_{b=T_1+1}^{T_1+T_2}\boldsymbol{1}\left(\varepsilon^{(b)}_{ij}>\xi_U(y_i)\right),\qquad\forall (i,j).
	\end{equation}
	At the end of the $T_2$ iterations, we determine the mixture $\widetilde{g}_\varepsilon$ that is effectively used as approximation:
	\begin{equation}\label{eq:mixt_choice}
		\widetilde{g}_\varepsilon=\left\{\begin{array}{l}
			g_\varepsilon \text{ for } \varepsilon_{ij} \text{ such that } \kappa_{ij}^U\leq p_U;\\
			g_\varepsilon^\star \text{ for } \varepsilon_{ij} \text{ such that } \kappa_{ij}^U>p_U.
		\end{array}\right.
	\end{equation}
	where the default value of $p_U$ is set to .05 in the \texttt{SamplerPoisson} package.
	\item \textit{Steps 1-2.} As IAMS algorithm.
	\item \textit{Step 3.} Draw the proposed value $\boldsymbol{\beta}^\text{prop}$ from the full conditional as \eqref{eq:FC_beta}, but using the adjusted mixture $\widetilde{g}_\varepsilon$. Accept the proposed value with probability 	
	\begin{equation}
		\pi\left(\boldsymbol{\beta}^{\text{prop}},\boldsymbol{\beta}^{(b-1)},\mathbf{y}^{*(b)}, \boldsymbol{\gamma}^{(b-1)}\right)=\frac{\mathcal{L}\left(\boldsymbol{\beta}^{\text{prop}}, \boldsymbol{\gamma}^{(b-1)}; \mathbf{y}^{*(b)}\right)}{\mathcal{L}\left(\boldsymbol{\beta}^{(b-1)}, \boldsymbol{\gamma}^{(b-1)}; \mathbf{y}^{*(b)}\right)}\cdot \frac{{\widetilde{\mathcal{L}}_a}\left(\boldsymbol{\beta}^{(b-1)}, \boldsymbol{\gamma}^{(b-1)}; \mathbf{y}^{*(b)}\right)}{{\widetilde{\mathcal{L}}_a}\left(\boldsymbol{\beta}^{\text{prop}}, \boldsymbol{\gamma}^{(b-1)}; \mathbf{y}^{*(b)}\right)};
	\end{equation}
	where $\widetilde{\mathcal{L}}_a\left(\boldsymbol{\beta}, \boldsymbol{\gamma}; \mathbf{y}^{*}\right)=\prod_{i=1}^{n}\prod_j\widetilde{g}_\varepsilon\left(y_{ij}^*-\mathbf{x}_i^\top\boldsymbol{\beta}-\mathbf{z}_i^\top\boldsymbol{\gamma}\right)$.
	\item \textit{Step 4.} Draw the proposed value $\boldsymbol{\gamma}_q^\text{prop},\ \forall q,$ from the full conditional as \eqref{eq:FC_gamma}, but using the adjusted mixture $\widetilde{g}_\varepsilon$. Accept the proposed value with probability 	
	\begin{equation}
		\begin{aligned}
			\pi\left(\boldsymbol{\gamma}_q^{\text{prop}},\boldsymbol{\gamma}_q^{(b-1)},\mathbf{y}^{*(b)},\boldsymbol{\beta}^{(b)}, \left\{\boldsymbol{\gamma}_{q^\prime}^\bullet\right\}_{q^\prime\neq q}\right)=&\frac{\mathcal{L}\left(\boldsymbol{\beta}^{(b)},\boldsymbol{\gamma}_q^{\text{prop}},\left\{\boldsymbol{\gamma}_{q^\prime}^\bullet\right\}_{q^\prime\neq q}; \mathbf{y}^{*(b)}\right)}{\mathcal{L}\left(\boldsymbol{\beta}^{(b)},\boldsymbol{\gamma}_q^{(b-1)},\left\{\boldsymbol{\gamma}_{q^\prime}^\bullet\right\}_{q^\prime\neq q}; \mathbf{y}^{*(b)}\right)}\times\\
			&\qquad\times\frac{\widetilde{\mathcal{L}}_a\left(\boldsymbol{\beta}^{(b)},\boldsymbol{\gamma}_q^{(b-1)},\left\{\boldsymbol{\gamma}_{q^\prime}^\bullet\right\}_{q^\prime\neq q}; \mathbf{y}^{*(b)}\right)}{\widetilde{\mathcal{L}}_a\left(\boldsymbol{\beta}^{(b)},\boldsymbol{\gamma}_q^{\text{prop}},\left\{\boldsymbol{\gamma}_{q^\prime}^\bullet\right\}_{q^\prime\neq q}; \mathbf{y}^{*(b)}\right)};
		\end{aligned}
	\end{equation}
	where $\boldsymbol{\gamma}_{q^\prime}^\bullet=\boldsymbol{\gamma}_{q^\prime}^{(b)}$ if $q^\prime<q$ and $\boldsymbol{\gamma}_{q^\prime}^\bullet=\boldsymbol{\gamma}_{q^\prime}^{(b-1)}$ if $q^\prime>q$.
	\item \textit{Step 5.} As IAMS.
\end{itemize}

For the sake of comparison, in the next section we will also consider the algorithm obtained by simply adding the MH step to the original IAMS algorithm, i.e. approximating the NLG distribution with the original mixture $g_\varepsilon$ instead of $\widetilde{g}_\varepsilon$. We label it as MH-IAMS algorithm and it follows the steps of the RIAMS algorithm but replacing $\widetilde{g}_\varepsilon$ and $\widetilde{\mathcal{L}}_a$ with ${g}_\varepsilon$ and ${\mathcal{L}}_a$.

We stress that the IAMS algorithm with mixture approximation \eqref{eq:mixt-app} is generally the best choice for sampling Poisson LGMs because of its computational efficiency, provided that the accuracy of the approximation does not compromise the convergence of the algorithm. In fact, the adjusted mixture $\widetilde{g}_\varepsilon$ is more computationally demanding than $g_\varepsilon$ because of the increased number of components and adding a rejection step is computationally demanding as it requires computation of the acceptance probability which implies computation of the exact and approximated likelihoods.

For these reasons, our ultimate goal is to implement an automatic algorithm that applies RIAMS or MH-IAMS only when the approximation could prevent IAMS from convergence. Recall that approximation issues arise when one or more residuals fall in the tails of the NLG distribution. If a residual lies in the left tail, the MH step can resolve the issue, while, if it falls in the right tail, the mixture must also be adjusted. As cut-off points for determining whether a residual falls in the distribution tails, we use the previously mentioned $\xi_U(y)$ for the upper tail and $\xi_L(y)$ for the lower tail. Both thresholds are defined as the points where the absolute value of the log difference between the NLG and mixture densities exceeds 1.

To put these considerations into practice, we introduce a preliminary step consisting of a training period to assess whether strengthening the IAMS algorithm is necessary. The preliminary step is structured as follows:
\begin{itemize}
	\item \textit{Step 0a.} Perform $T_1$ iterations with the IAMS algorithm. 
	\item \textit{Step 0b.}	Perform $T_2$ iterations and compute, as in the RIAMS algorithm, the proportions defined in \eqref{eq:upper}. In addition, we also compute, for each residual, the proportion of iterations exceeding the lower bound:
	\begin{equation}
		\label{eq:condition_mh}
		\kappa_{ij}^L=\frac{1}{T_2}\sum_{b=T_1+1}^{T_1+T_2}\boldsymbol{1}\left(\varepsilon^{(b)}_{ij}<\xi_L(y_i)\right),\qquad\forall (i,j).
	\end{equation}
	This corresponds to checking if for some observation the generated residuals are in a region where the mixture approximation \eqref{eq:mixt-app} is inaccurate. At the end of the $T_2$ iterations the effectively used mixtures are selected as in \eqref{eq:mixt_choice} and the algorithm to use for the subsequent iterations is chosen according to the following rule:
	\begin{itemize}
		\item IAMS if $\kappa_{ij}^L<p_L$ and $ \kappa_{ij}^U<p_U,\ \forall (i,j)$;
		\item MH-IAMS if $\exists (i,j) \text{ s.t. } \kappa_{ij}^L>p_L,$ and $ \kappa_{ij}^U<p_U,\ \forall (i,j)$;
		\item RIAMS if $\exists (i,j) \text{ s.t. } \kappa_{ij}^U>p_U$. Note that, in this case, the adjusted mixture $\widetilde{g}_\varepsilon$ is adopted only on the residuals that are individuated as problematic.
	\end{itemize}
	The choice depends on two thresholds $p_L$ and $p_U$ that we advise to set equal to small values. In the \texttt{SamplerPoisson} package, the default value is $p_L=p_U=0.05$.
\end{itemize}

The Automatic sampler, as well as all samplers discussed in this Section, are implemented in the \texttt{SamplerPoisson} package available in the Supplementary Material.

\section{Applications}\label{sec:applications}

In this Section, we present two applications to demonstrate the usefulness of the algorithms proposed in Section \ref{sec:mh-iams-auto}. The first application generalizes the toy example introduced in Section \ref{sec:pitfalls}, while the second uses a real dataset from \cite{alain2013}, which details squirrel behavior and forest attributes across various plots in Scotland's Abernathy Forest. For both examples, we use the MCMC algorithm implemented in Stan as a benchmark. All computations are carried out using the \texttt{SamplerPoisson} R package, and the R code for estimation and result summaries is provided as supplementary material.

\subsection{Toy example: reprise}

\begin{figure}
	\centering
	\includegraphics[width=1\linewidth]{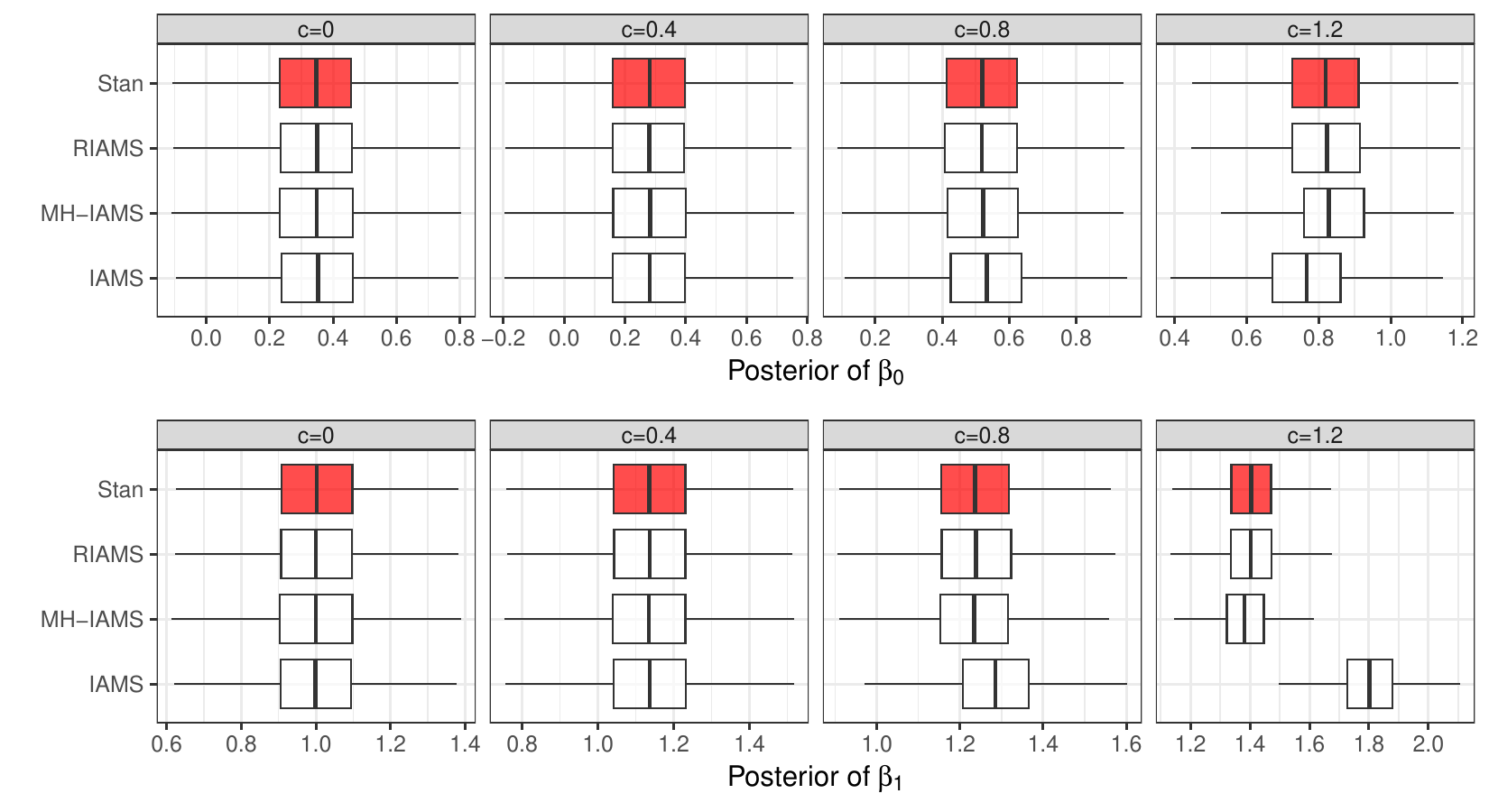}
	\caption{Posterior distributions of model parameters. Each panel shows the posterior distribution of one parameter for a given $c$ value (reported in the title) obtained with the considered algorithms.}
	\label{fig:toy_all}
\end{figure}

In this section, we show results obtained by using our proposed algorithm in an extension of the toy example introduced in Section \ref{sec:pitfalls}.
In particular, four datasets are generated from a Poisson distribution with intensity parameter
\begin{equation*}
	\log\left(\lambda_i\right)=0.1+x_{1i} + c x_{2i},\qquad i = 1,\ldots,30,
\end{equation*}
where both covariates are drawn from a standard Gaussian distribution and $c\in\{0,0.4,0.8,1.2\}$. Subsequently, models are estimated omitting the covariate $x_2$ to simulate increasing levels of model misspecification. We then compare the performance of the IAMS algorithm with that of the algorithms described in Section \ref{sec:mh-iams-auto}. Specifically, the comparison between the RIAMS and MH-IAMS algorithms highlights cases where the adjusted mixture approximation enhances convergence. Finally, the Automatic algorithm is evaluated to assess its ability to adaptively switch to a more robust method when necessary. For each algorithm we run a chain of 110,000 iterations, discarding the first 10,000 iterations as burn-in. For the Automatic algorithm, we set $T_1=500$, $T_2=250$ and $p_L=p_U=0.05$.

Figure \ref{fig:toy_all} presents the posterior distribution of $\beta_0$ and $\beta_1$ for the different values of the omitted variable coefficient $c$, across the MCMC algorithms under study. As shown in the plots, all algorithms converge to the target distributions when $c=0$ and $c=0.4$. However, for $c=0.8$ and $c=1.2$, the posterior distributions obtained using the IAMS algorithm deviate from the Stan benchmark for both the intercept $\beta_0$ and the regression coefficient $\beta_1$. Furthermore, the MH-IAMS algorithm also exhibits discrepancies from the target posterior when $c=1.2$, witnessing that the mixture adjustment in the right tail, introduced in Section \ref{sec:mh-iams-auto}, can improve convergence. In contrast, the RIAMS sampler remains in  agreement with the target posterior across all values of $c$, providing evidence in favor of its robustness in presence of extreme residuals generated by the omission of a relevant covariate.

\begin{table}[]\centering
	\begin{tabular}{@{}lcccc@{}}
		\toprule
		Algorithm  & $c=0$ & $c=0.4$ & $c=0.8$ & $c=1.2$ \\  \midrule
		MH-IAMS    & 1.00 & 1.00 & 0.72 & 0.02 \\
		RIAMS& 1.00 & 1.00 & 0.87 & 0.74 \\ 
		Automatic       & IAMS & IAMS & 0.87 & 0.75 \\ \bottomrule
	\end{tabular}
	\caption{Acceptance rates for each algorithm as a function of the underlying coefficient $b$.}
	\label{tab:acc}
\end{table}

\begin{table}
	\centering
	\begin{tabular}{@{}lcccc@{}}
		\toprule
		Algorithm  & $c=0$ & $c=0.4$ & $c=0.8$ & $c=1.2$ \\  \midrule
		MH-IAMS & 2.43 & 2.40 & 2.46 & 2.01 \\ 
		RIAMS & 2.49 & 2.38 & 2.62 & 2.38 \\ 
		Automatic & 1.00 & 1.00 & 2.65 & 2.01 \\  \bottomrule
	\end{tabular}
	\begin{tabular}{lcccc}
	\end{tabular}
	\caption{Relative computational times with respect to IAMS with 50 replications, 10,000 MCMC iterations each. Average elapsed times for Gibbs samplers ranges from 31$s$ to 36$s$.}
	\label{tab:toy_times}
\end{table}

These results are closely linked to the acceptance rates observed for the compared algorithms, which are reported in Table \ref{tab:acc} as a function of $c$. First, we note that the acceptance rate of the RIAMS algorithm is higher than that of the MH-IAMS algorithm for $c=0.8$ (0.87 vs 0.72). Furthermore, when $c=1.2$, the acceptance rate of MH-IAMS drops significantly to 0.01, leading to poor mixing, whereas the RIAMS algorithm continues to perform satisfactorily. Regarding the Automatic sampler, it can be seen that the IAMS algorithm is selected when $c=0$ and $c=0.4$, which is an appropriate choice since, in these cases, the IAMS algorithm converges, and methods involving an MH step achieve an acceptance rate of 1. Conversely, for $c=0.8$ and $c=1.2$, the Automatic sampler correctly switches to the RIAMS algorithm, as the IAMS algorithm struggles to converge due to the presence of residuals in the NLG tails.

The ability of the Automatic sampler to select the RIAMS algorithm only when necessary proves particularly advantageous when considering computational time. Table \ref{tab:toy_times} reports the ratio of the computational time of each algorithm to that of the IAMS algorithm. The results show that algorithms incorporating the MH step require more than twice the computational time of the IAMS algorithm. The key advantage of the Automatic sampler is that it applies this additional computational effort only when the IAMS algorithm risks failing to converge due to inaccuracy of the mixture approximation.

\subsection{Nuts data}

In this Section we deal with an application of Poisson regression to the \texttt{nuts} dataset available in the R-package \texttt{COUNT} adopted in \cite{alain2013}. The dataset comprises observations on $n=52$ plots providing information about squirrel behavior and forest attributes across various plots in Scotland's Abernathy Forest. The application in what follows is not meant to provide a sound statistical analyses of these data but to illustrate the merit of the algorithms proposed in Section \ref{sec:mh-iams-auto}. The response variable represents the number of cones stripped by squirrels. The linear predictor is specified as the sum of two linear effects, i.e. the standardized mean tree height ($\mathbf{x}_\text{height}$) and standardized canopy closure per plot ($\mathbf{x}_\text{canopy}$), plus a smooth effect of the standardized number of trees per plot ($\mathbf{x}_\text{trees}$). 

\begin{figure}
	\centering
	\includegraphics[width=1\linewidth]{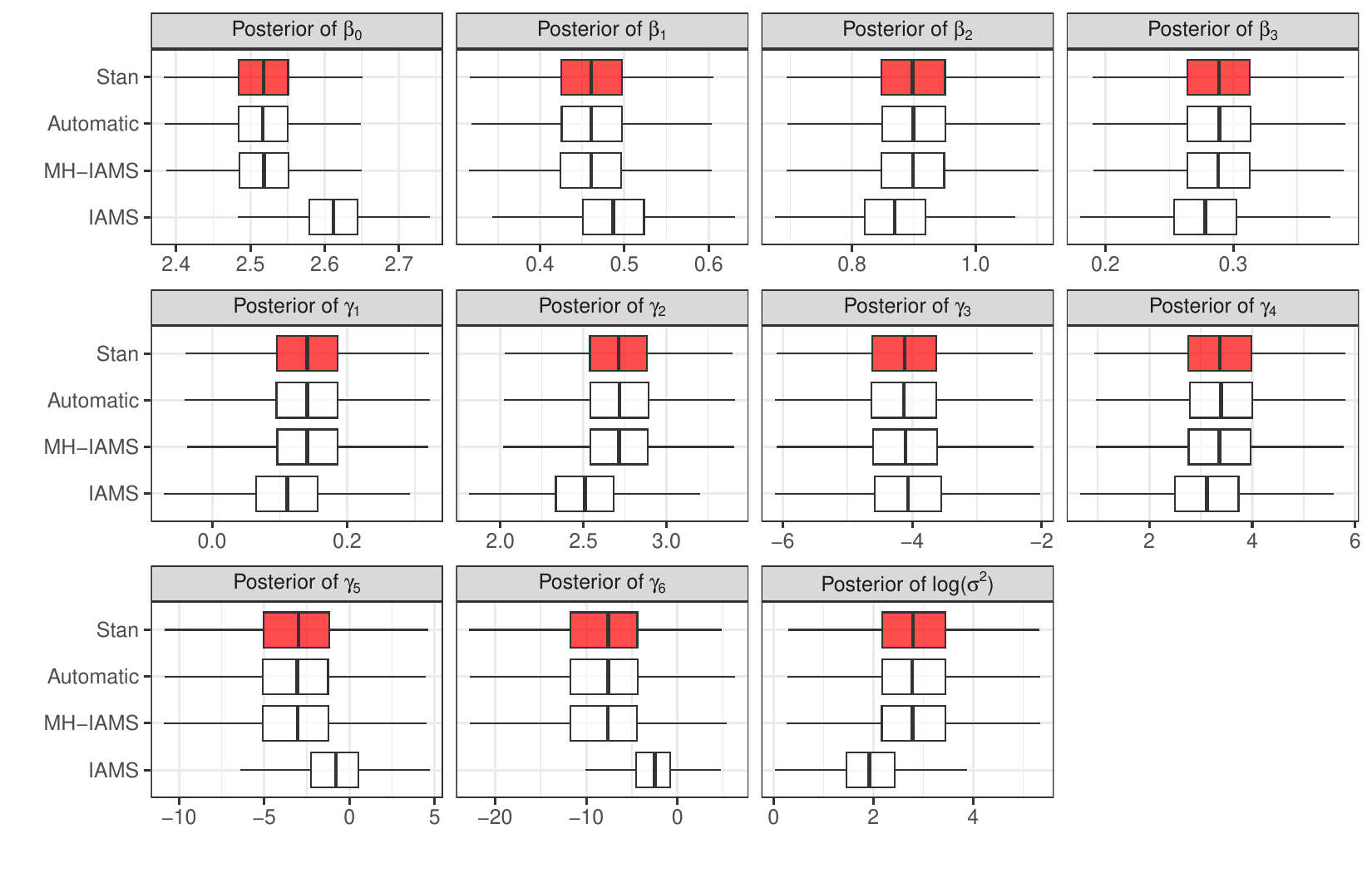}
	\caption{Posterior distributions of model parameters. Each panel shows the posterior distributions of one parameter obtained under the considered algorithms.}
	\label{fig:nuts1}
\end{figure}

This specification delivers a Latent Gaussian Model where the likelihood is
\[
y_{i}|\lambda_i\sim\mathcal{P}(\lambda_i),\qquad i=1,\ldots,52,
\]
and the vector of linear predictors is specified as
$$
\log(\boldsymbol{\lambda})=\beta_0+\beta_1\mathbf{x}_\text{height}+\beta_2\mathbf{x}_\text{canopy}+f(\mathbf{x}_\text{trees}).
$$
The term $f(\mathbf{x}_\text{trees})$ indicates the assumption of a smooth relationship between $\mathbf{x}_\text{trees}$ and the linear predictor, that we implement through the popular Bayesian P-splines \citep{lang2004bayesian} as:
\begin{equation*}
	f(\mathbf{x}_\text{trees})=\mathbf{Z}\boldsymbol{\delta},\quad \boldsymbol{\delta}|\omega^2\sim\mathcal{N}_m\left(\boldsymbol{0}, \omega^2\mathbf{K}^-\right),
\end{equation*}
where $\mathbf{Z}\in\mathbb{R}^{n\times m}$ is made by cubic B-spline basis functions evaluated at $m$ knots, while $\boldsymbol{\delta}$ is the vector of coefficients for which we assume a second-order random walk prior, determined by the rank-deficient precision matrix $\mathbf{K}$. To 
get model specification that is easily portable also in \texttt{Stan}, we opt for the generic mixed model parameterization described in \citet{scheipl2012spike}:
\begin{equation*}
 f(\mathbf{x}_\text{trees})=\beta_3\mathbf{x}_\text{trees}+\mathbf{Z}_\text{o}\boldsymbol{\gamma},\quad \boldsymbol{\gamma}|\sigma^2\sim\mathcal{N}_{m^\prime}\left(\boldsymbol{0},\sigma^2\mathbf{I}_{m^\prime}\right);
\end{equation*}
where $\mathbf{Z}_\text{o}\in\mathbb{R}^{n\times m^\prime}$ is a matrix of $m^\prime<m$ orthonormal basis obtained exploiting the spectral decomposition of the covariance matrix of $\mathbf{Z}\boldsymbol{\delta}$, i.e. $\mathbf{ZK}^-\mathbf{Z}^\top$, which has rank $m^\prime$.
Model hierarchy is completed by prior specification for fixed effect coefficients $\beta_k\sim\mathcal{N}(0,1000)$, $k=0,1,2,3$ and for the scale parameter $\sigma^2$ for which we chose $\sigma^2\sim\text{Gamma}(1, 0.001)$.

\begin{table}[]
	\centering
	\begin{tabular}{@{}lrrrr@{}}
		\toprule
		& \multicolumn{2}{c}{MH-IAMS}          & \multicolumn{2}{c}{RIAMS}        \\ \cmidrule(l){2-3} \cmidrule(l){4-5} 
		Parameter     & $n_\text{eff}$ & $n_\text{eff}/\text{sec}$ & $n_\text{eff}$ & $n_\text{eff}/\text{sec}$ \\ \midrule
		$\beta_0$   & 1429           & 30                  & 4283           & 77                  \\
		$\beta_1$   & 2365           & 49                  & 7208           & 129                 \\
		$\beta_2$  & 1731           & 36                  & 4474           & 80                  \\
		$\beta_3$  & 2943           & 61                  & 8839           & 159                 \\\midrule
		$\gamma_1$ & 6529           & 136                 & 8789           & 158                 \\
		$\gamma_2$ & 7580           & 158                 & 13604          & 244                 \\
		$\gamma_3$ & 4613           & 96                  & 6793           & 122                 \\
		$\gamma_4$ & 8899           & 186                 & 9197           & 165                 \\
		$\gamma_5$ & 6339           & 132                 & 9122           & 164                 \\
		$\gamma_6$ & 3798           & 79                  & 5464           & 98                  \\\midrule
		$\sigma^2$  & 6964           & 145                 & 10327          & 185                 \\ \bottomrule
	\end{tabular}
	\caption{Effective sample size ($n_{\text{eff}}$) and effective sample size per second ($n_{\text{eff}}/\text{sec}$) for the MH-IAMS and RIAMS algorithms, $B=100,000$ MCMC iterations.}
	\label{tab:neff-nuts}
\end{table}

\begin{table}[]
	\centering
	\begin{tabular}{@{}lrrr@{}}
		\toprule
		& IAMS  & MH-IAMS & RIAMS \\ \midrule
		Acceptance rate for $\boldsymbol{\beta}$  & -     & 0.23    & 0.62      \\
		Acceptance rate for $\boldsymbol{\gamma}$ & -     & 0.58    & 0.76      \\
		Elapsed time (for $B=100,000$)  & 13$s$ & 47$s$   & 56$s$     \\ \bottomrule
	\end{tabular}
	\caption{Acceptance rates and elapsed time of the compared algorithms.}
	\label{tab:accrates}
\end{table}

In what follows, we compare results obtained by implementing the MCMC sampler in Stan, which again serves as a benchmark, with results obtained by the IAMS, MH-IAMS, and Automatic samplers, which in this application selects the RIAMS algorithm. Indeed, the preliminary checks pointed out that among the $99$ latent variables introduced by the first augmentation scheme, $\kappa_{ij}^L$ and $\kappa_{ij}^U$ defined in Equations \eqref{eq:upper} and \eqref{eq:condition_mh} exceeded the threshold $0.05$ in $9$ and $3$ cases, respectively. Figure \ref{fig:nuts1} reports the posterior distribution of each model parameter for the compared algorithms. It can be noticed that posterior distributions obtained with the IAMS algorithm do not overlap with Stan posteriors for all model parameters, highlighting a lack of convergence. On the other hand, the Automatic (RIAMS) and the MH-IAMS algorithms show good agreement with Stan posteriors: hence, adding a rejection step is needed in order to achieve convergence in this application.

Table \ref{tab:neff-nuts} compares the performance of the MH-IAMS and Automatic algorithms, reporting the effective sample size and the effective sample size per second for each, based on $B=100,000$ MCMC iterations after a burn-in of 10,000 iterations. The results show that the RIAMS algorithm, correctly chosen by the Automatic procedure, achieves significantly higher effective sample sizes, both in absolute terms and relative to elapsed time.This improvement is closely linked to the adjustment of the mixture in the right tail for problematic residuals, which leads to a significant increase in acceptance rates under the RIAMS algorithm. As shown in Table \ref{tab:accrates}, the acceptance rate for $\boldsymbol{\beta}$ increases substantially (from 0.23 to 0.62), while $\boldsymbol{\gamma}$ sees a moderate rise (from 0.58 to 0.76). This directly translates into a reduction in chain autocorrelation.
Finally, regarding total computational time, the IAMS algorithm is significantly faster (13 seconds for 110,000 iterations), as the necessary additional rejection step increases computational cost (47 seconds for the MH-IAMS algorithm and 56 seconds for the Automatic sampler).

\section{Concluding remarks}\label{sec:conclusion}
The IAMS algorithm is a particularly appealing and computationally efficient method for sampling from the posterior distributions of Poisson LGMs. It enables the use of standard sampling algorithms originally designed for LGMs with Gaussian likelihoods, once auxiliary variables have been generated. However, this widely used auxiliary mixture sampling approach relies on an approximation step, whose impact has yet to be fully explored in the literature.

The IAMS algorithm usually provides an accurate approximation of the posterior distribution in Poisson regression models. However, in certain applications where extreme residuals are observed relative to the augmented model, the algorithm may fails to converge to the true posterior distribution. This is primarily due to the rapid decay of the right tail in the adopted mixture approximation, as discussed in Section \ref{sec:pitfalls}. It is important to note that the failure of the approximation can be particularly insidious, as there are no explicit warnings to alert users. A useful first step is to compare the true log-likelihood with the approximated one. This check is generally causes limited computational burden, and any observed discrepancy can serve as an indicator of potential issues with the approximation.

In this paper, we proposed a robust version of the IAMS algorithm, involving an acceptance step during the sampling process and an adjusted mixture approximation that provides a more accurate approximation of the right tail of the NLG distribution, favoring convergence to the posterior distribution even in presence of extreme residuals. Since both the acceptance step and the employment of the adjusted mixture require additional computational effort with respect to the IAMS algorithm, a sampler that judge for the need of such algorithm during a training period has been designed and implemented in the R package \texttt{SamplerPoisson}.

\bibliographystyle{abbrvnat}
\bibliography{sample.bib}

\end{document}